# PERSONALISED PRODUCT DESIGN USING VIRTUAL INTERACTIVE TECHNIQUES

Kurien Zacharia[1], Eldo P. Elias[2], Surekha Mariam Varghese[3]

Department of Computer Science and Engineering, M. A. College of Engineering, Kothamangalam, Kerala, India
`kurienzach@gmail.com; eldope@gmail.com; surekha@mace.ac.in`

## ABSTRACT

*Use of Virtual Interactive Techniques for personalized product design is described in this paper. Usually products are designed and built by considering general usage patterns and Prototyping is used to mimic the static or working behaviour of an actual product before manufacturing the product. The user does not have any control on the design of the product. Personalized design postpones design to a later stage. It allows for personalized selection of individual components by the user. This is implemented by displaying the individual components over a physical model constructed using Cardboard or Thermocol in the actual size and shape of the original product. The components of the equipment or product such as screen, buttons etc. are then projected using a projector connected to the computer into the physical model. Users can interact with the prototype like the original working equipment and they can select, shape, position the individual components displayed on the interaction panel using simple hand gestures. Computer Vision techniques as well as sound processing techniques are used to detect and recognize the user gestures captured using a web camera and microphone.*

## 1. INTRODUCTION

Customized products are great ways to market a particular company. For large companies manufacturing of custom products is not a problem because of the flexibility to run large and small production runs for components. In contrast, small companies or individuals need proper strategic planning to have such personalized products. In the current scenario, the electronics industry does not provide much facility to incorporate customer specific changes to the existing models and it is difficult to incorporate specific changes such as change in the basic features or inserting special applications to products such as mobile phones according to the customer interest.

Many different prototyping methods are available to suit the method of manufacturing of the product. Physical mockups (Hardware Prototyping) have always played an important role in the early conceptual stages of design. It involves creating a static model of the product from cheap and readily available materials such as paper, cardboard, Thermocol etc. They are commonly used in the design and development electronic equipments such as mobile phones. Physical mockups are used in early stages of design to get a feel of the size, shape and appearance of the equipment.

              1

Another kind of prototyping that is particularly used for design of electronic equipment's is the software simulation. The software required for the equipment is written and then run in simulator software inside the computer. This method allows testing the software of the product and the user interactions with it.

The disadvantage of the above methods is that the user experience remains disjoint. In Hardware prototyping the user gets the touch and feel of the product but does not include the working or interactions with the product. In software based simulations the user gets to know how the product would interact but he has no touch and feel of the actual product. Other methods of prototyping like using a prototyping workbench allows to create hardware interface with relative ease, but the cost factor makes them beyond the reach of most small and medium scale industries.

None of the previously mentioned methods yield personalized products. This paper describes Personalized Product Design (PPD) using the technique Virtual Interactive Prototyping (VIP) described in [1]. VIP is a virtual reality prototyping method based on Sixth Sense[2] and an existing prototyping method called Display Objects[3].

PPD allows selecting and organizing previously created components easily by using computer simulated objects displayed on to an actual physical model using a projector. The user can create a model of his choice on the computer and test it out using a real physical model without actually assembling any component using the concept of augmented reality. Physical models of the actual product in different shape and size are constructed using Paper, Cardboard or Thermocol. The components of the equipment such as (viz. screen, buttons etc.) are then projected from a projector connected to the computer into the physical model. A web camera tracks the user's gestures as well as the physical model and reacts appropriately. The accuracy of touch detection using computer vision alone is not sufficient for sensing. Therefore a microphone is attached to the surface, and the change in input level from the touch of the finger is detected. By combining the input from the microphone and camera a usable interactive system is developed. PPD uses Computer Vision techniques as well as Sound processing techniques to detect user interactions so that the user can touch and interact with the prototype like actual working equipment. The system model used in PPD is shown in Figure 1.

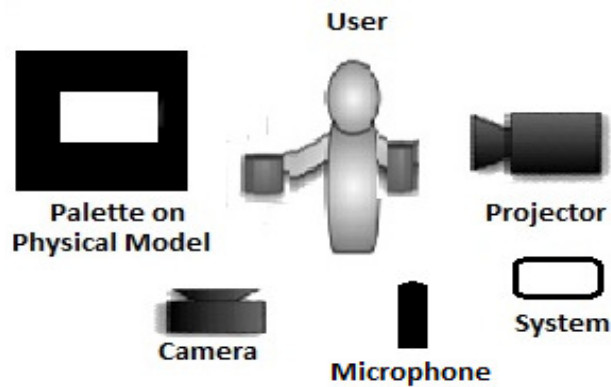

Figure 1. The Model

## 2. RELATED WORK

Display objects [3] introduced the concept of virtual models that can be designed on a Macintosh system and then projected to real physical models called display objects. The disadvantage of the system is that user interactions in this system is based on expensive high end infrared based camera which require calibration each time the system is put to work.

Sixth Sense is an innovative computing method that uses user gestures to perform day to day activities. Sixth Sense uses a camera, a mini projector, microphone, and color markers to track fingers. Sixth Sense suggests a new method of interaction. It is used for a variety of interesting applications such as checking flight delays with flight ticket and automatically getting the reviews of books just by showing its Cover etc. In Sixth sense, markers with distinct color are placed in each of the fingers. Positions of the fingers are identified by locating the positions of the color markers.

The concept of interactive techniques for product design has been used elsewhere. Avrahami, D. and Hudson used interactive techniques for product design [4]. Use of static virtual artefacts has been proposed for the analysis of team collaboration in virtual environments [5]. The display method uses projections to render a realistic model of the object according to the users viewing position. Maldovan et. al used a virtual mock-up process for reviewing designs of buildings and civil structures such as court rooms, restaurants[6]. Pushpendra et.al. proposed an approach using immersive video with surround sound and a simulated infrastructure to create a realistic simulation of a ubiquitous environment in the software design and development office[7].

Gesture based interaction techniques are being used for interaction in many innovative designs. [8] presents a vision-based interface for instructing a mobile robot through both pose and motion gestures. A declarative, model based gesture navigation design for rapid generation of different prototypes is described in [9].

## 3 MODELLING THE PRODUCT

To construct a prototype the user will have to first drag and drop all the necessary components from a virtual palette into the virtual prototype using hand gestures. The components can be then rearranged, moved and resized inside the Virtual Prototype. After all the required components are set up, the user can then start the software simulation of the product. This means he can use it like real equipment, in the way the programmer has programmed it. The user can have both the touch and feel of a real hardware prototype as well as all the interactions.

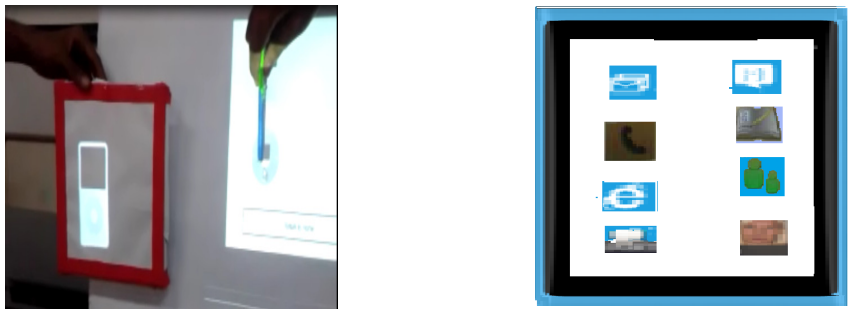

Figure 2. Phone in Display Object

## 4  INTERACTION TECHNIQUES

The workbench incorporates both one-handed and two handed techniques so that the user can interact with the model with one hand while browsing interaction elements with the other. The user can interact with the model with simple hand gestures. With one hand the user can hold the physical model on which the virtual model is displayed which is named as Display Object. Interaction elements are then copied and pasted onto the display object using simple gestures of fingers augmented with color marker. The important hand gestures such as Scanning, Selecting, Placing, Dragging, Locking, Clicking, Resizing, Wiping are described in [10]

## 5  COMPONENT PROGRAMMING

In component programming what are the icons, how each one should look like and where each to be placed and what are the functionalities associated with the icons are decided by the user. The user will be provided with lists of icons and functionalities. He can either select from the existing lists or can create new icons and functionalities.

After display elements are placed on the physical model, the panel that can be edited in one of three ways: on a computer, on the Palette, or on the physical model itself. Size and position of the icons can be edited using the edit menu on the palette or on the Display object. Using hand gestures and with the help of sound tracking icons can be edited on the Display object. The physical model is placed in edit mode by pressing a button on the Palette, or by providing a wipe gesture on its surface. There are two types of elements on the palette: input elements such as buttons and output elements such as display. Different action sequences are programmed and maintained as components. The users can connect elements with functionalities/actions. For example a particular user can select from a list of display functions and can associate a screen element with the particular display. A movie element, in turn, can be scripted through the contents of a movie file connected to its action. Simple interactive behaviours are limited to simple actions, such as starting, stopping or scrolling through the movie file displayed on the screen element. More complicated behaviours can be made possible by combining different actions. Since display panel is made out of real materials, it is easy to extend its behaviour with real world artefacts as well, mixing properties of bits with those of atoms. For example, paper sketches or physical buttons can be easily affixed on the physical prototype, and linked with interactive content. This not only allows quick iterative revisions of the physical model, but also allows for physical elements to be mixed with digital elements, for example, to provide feedback for an on-screen input element. One example of this is the use of physical pushpins to simulate surface effects of buttons displayed on the surface. This allows for feedback when interacting with the virtual prototype.

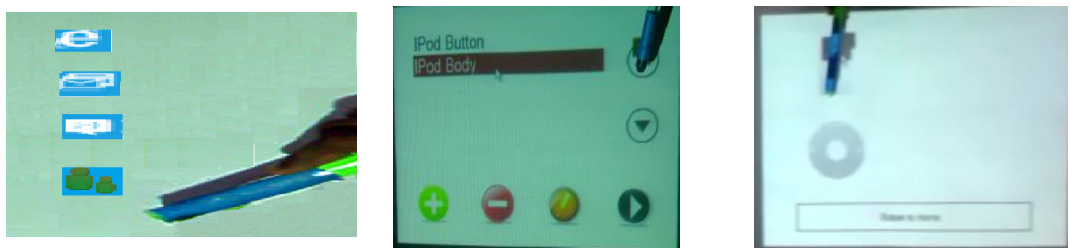

Figure 3. Designing the personalized mobile phone

International Journal of Computer Graphics & Animation (IJCGA) Vol.2, No.1, January 2012

## 6 IMPLEMENTATION

Common Image and sound processing techniques are extensively used in implementing the concept. EmguCV, which is a C# wrapper for the OpenCV image processing library developed by Intel is used for image processing. The Open Source Computer Vision Library (OpenCV)[11] is a comprehensive computer vision library and machine learning written in C++ and C with additional Python and Java interfaces. It officially supports Linux, Mac OS, Windows, Android etc.

The programming language used is C#.net as it facilitates fast development. For audio capturing and processing BASS audio processing library is used. All the core libraries that are used for implementing VIP are open source and platform independent. However VIP is implemented in Windows platform.

Personalized Product Design is implemented with the help of 3 modules. Marker Object tracking, Display Object Tracking and Object Rendering. The first two modules deal with object tracking. Fingers capped with color markers are identified, located and traced in the marker Object Tracking Module. The virtual model or the projection of the model on the physical mockup can be moved within the permissible limits. The movement of the model is tracked using the module Display Object Tracking. Marker object tracking is based on color segmentation. Color segmentation approach is not suitable for Display object tracking Object, since colors are not significant in the display object. Hence edge detection using Canny algorithm is used for identifying the Display object. Rendering module deals with the rendering of the product image on the physical model. The details of implementation are discussed in [10].

The major processing modules are

### 6.1 Image Processing

User interacts with the model using hand gestures. Color markers are placed on the user's fingers for distinguishing the commands. For tracking the interactions of the user, the position and colour of the fingers are sensed using standard image processing algorithms for edge detection and segmentation.

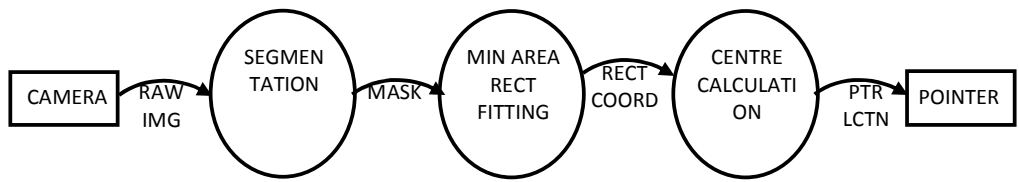

Figure 4. Level-1 DFD for Camera Capture

An HSV Color space based Color segmentation algorithm is used to isolate the color of the color marker. The centre point of the segmented region is roughly taken as the position of the user's finger. For tracking the motion of the display object, first corners of the physical prototype are located. A greyscale based edge detection algorithm is used for this purpose. The image of the product is then fitted into this rectangle using image transforms.

**Color Image Segmentation**

For tracking the user actions, the image of the color marker should be separated from the picture captured by the live camera. This process of separating out mutually exclusive homogeneous regions of interest is known as color image segmentation. Image segmentation has been the subject of considerable research activity over the last three decades. Many algorithms have been elaborated for this purpose [12][13][14].

The segmentation algorithm removes all other color from the input image other than the color of the color marker. Output of the segmentation algorithm is a black and white image with all areas black except the portions having the color of the color marker. The features extracted from this are the edges. They are extracted by looking for the difference between pixel values. Where there is an abrupt change in the pixel value (such as between black and white pixels) it is defined as an edge.

The segmentation is done in five steps, Filtering, Formation of image, Approximate object area determination, Object identification and Final Segmentation. Smoothing filters are, in general, used for noise removal and blurring. Blurring is used as a preprocessing step to remove small details from the image prior to extraction of large objects as well as bridging of small gaps in lines and curves. Here Gaussian filter is used just before the image segmentation step to smooth out the image. This removes hard corners of the image and improves efficiency of the noise reduction process. Pyramid Up and Pyramid Down functions in the OpenCV library are used to remove noise from the binary mask images, which consist of a black background and white tracked object. The Pyramid Up and Pyramid Down methods reduce white noise pixels. The Pyramid Up expands the black pixels to the required number of pixels so that small white pixels if any will be converted to black pixels. This will reduce the size of white pixels which are valid. Pyramid Up process is invoked after this to restore the original size of white pixels by expanding the white pixels by the specified number of pixels.

Color values are represented in 8 bit format in the system. An HSV based color segmentation algorithm is used for better efficiency. Images are represented in 8 bit bitmapped format, which used a 3 dimensional matrix for storing pixel values. The threshold values for Hue, Saturation and Value are represented as HueThreshL, HueThreshH, SatThreshL, SatThreshH, ValThreshL, ValThreshH. Algorithm 2 describes the steps in the segmentation.

## Algorithm 1 – Segmentation Algorithm

```
for i=0 to imgmat->width
for j=0 to imagmat->height
color<-imgmat[x,y];
if(color.hue < HueThreshH && color.hue >HueThreshL &&
color.sat>SatThreshL && color.sat <SatThreshH &&
color.Value > ValThreshL && color.Value < ValThreshH)
imgmat[x,y] = 255;
else imgmat[x,y] = 0;
```

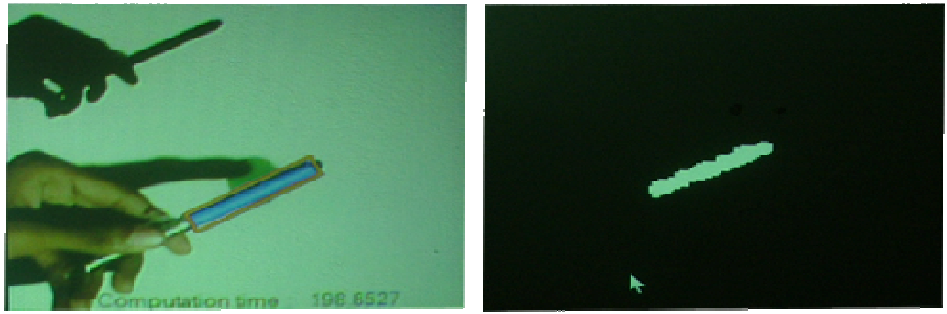

Figure 5. Image segmentation applied to marker object

**Edge Detection**

Color image segmentation techniques based on edge detection is a promising area of research [15]. In a noisy environment, canny algorithm is suitable for edge detection [16][17]. Canny's algorithm uses the gradient of the pixel to determine the edges of the Display object. Edge detection based on Canny's algorithm is faster than color segmentation for object identification. A pixel is considered as an edge pixel only if the gradient is above the upper threshold and is rejected if gradient is less than the lower threshold. If the pixel's gradient is between the thresholds, then the acceptance of the pixel is based on the direct connectivity of the pixel with pixels having gradient above the high threshold.

### Algorithm 2- Edge Detection Algorithm

```
for i=0 to imgmask.width
for j=0 to imgmask.height
for k=0 to 8
if (color[imgmask[x,y]-imgmask[x,k%3]]>colorthreshold)
if(x>maxx && y > maxy)
   point3 = point(x,y);
else if(x>maxx && y < miny)
   point2 = point(x,y);
else If(x<minx && y> maxy)
   point4 = point(x,y);
  else if(x<minx &y<miny)
   point1 = point(x,y)
```

**Object Tracking**

To find out the object, convex hull based algorithm is applied to the edge image. Boundary of the object is defined by the convex hull of the set of edge pixels. The features of the convex hull are extracted and stored in a feature list. The tracking algorithm works based on the extracted features. The features are tracked for different kinds of interactions. The input to the tracking algorithm is the Binary mask produced by the segmentation algorithm. The tracking of objects is done by analysing the centre positions of the objects that are tracked. The movement of the object is considered as a valid move only if the centre point of the current bin mask and previous bin mask has a difference of at least 8 pixels. Object Tracking algorithm is described in Algorithm 3.

**Algorithm 3- Object Tracking Algorithm**

```
dx = (currentimage.centre.x – previousimage.centre.x)
dy = (currentimage.centre.y – previousimage.centre.y)

if (dx >= 8 || dy >= 8)
{
    centre.x = currentimage.centre.x;
    centre.y = currentimage.centre.y;

}
```

### 6.2 Sound Processing

Image processing techniques have limitations in detecting whether the user has touched the virtual prototype or not. Therefore we use a microphone attached to the Virtual Prototype to detect whether the user has touched it or not. The input from the microphone is first passed through a Band Pass Filter to reduce background noise and then the amplitude of the resultant signal is compared with a threshold value to detect taps.

## 7. LIMITATIONS

The system is sensitive to lighting conditions, since it mainly rely on image capturing and processing. Too bright or too low light can hamper the system performance. Color segmentation cannot be accurately done when the background is fast moving.

## 8. CONCLUSION

A key issue in the new electronic era is the effort required for rapid prototyping and evaluation of user interfaces and applications. PPD provides a low-cost and rapid means to incorporate personalization of user interfaces and applications for custom-made products. Personalisation makes it easier for the user to use the required services. It enhances simplicity. Personalized systems will help in defining, maintaining and expressing the identity of the user.

PPD make use of the available information by means of already existing components using hand gestures. This enables reuse of components and reduces development time and gives better user satisfaction. Only necessary components are included in the actual product. This saves the electronic equipments used in prototyping and reduces electronic waste contributing to environment preservation and overall sustainable development. PPD can be termed as Green Technology.

PPD is very suitable for the new era of component based development. Components suitable for a particular user can be selected according to the particular user's need and taste. PPD can also be used in the design of custom-fit or tailor-made products. A customized product would imply the modification of some of its characteristics according to the customer's requirements such as with in the components of the particular mobile. However for custom-fit products, customization could be both in terms of the geometric characteristics of the body and the individual customer requirements. Dreaming for a future with tailor-made products where the users can select from a supermarket of components to insert, delete and modify them as necessary.

## AUTHORS

Kurien Zacharia completed his B-Tech in Computer Science and Engineering from M.A. College of Engineering, Kothamangalam in 2011. He has presented many papers in national level competitions and has published 2 papers in international journals and conference proceedings. His areas of research are Data Structures, In-memory databases, Computer Graphics and Virtual Reality.

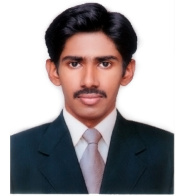

Eldo P Elias completed his BE in Computer Science and Engineering from Sri Ramakrishna Engineering College, Coimbatore in 2003. He has more than 7 years of teaching experience at M.A College of Engineering, Kothamangalam. He has published 2 papers in international journals and conference proceedings. His areas of research are Networking, Operating Systems and Computer Security.

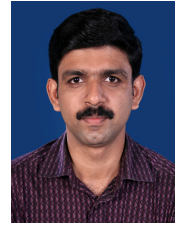

Surekha Mariam Varghese is currently heading the Department of Computer Science and Engineering, M.A. College of Engineering, Kothamangalam, Kerala, India. She received her B-Tech Degree in Computer Science and Engineering in 1990 from College of Engineering, Trivandrum affiliated to Kerala University and M-Tech in Computer and Information Sciences from Cochin University of Science and Technology, Kochi in 1996. She obtained Ph.D in Computer Security from Cochin University of Science and Technology, Kochi in 2009. She has around 20 years of teaching and research experience in various institutions in India. Her research interests include Network Security, Database Management, Data Structures and Algorithms, Operating Systems and Distributed Computing. She has published 8 papers in international journals and international conference proceedings. She has served as reviewer, committee member and session chair for many international conferences and journals.

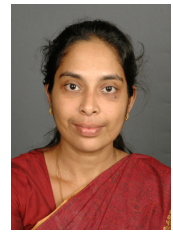